\documentclass[prl, aps, twocolumn, floatfix, showpacs]{revtex4}
\usepackage{graphicx, amsmath, amssymb, times}

\begin{document}
\title{Stability of spin-orbit coupled Fermi gases with population imbalance}
\author{M. Iskin$^1$ and A. L. Suba{\c s}{\i}$^2$}
\affiliation{
$^1$Department of Physics, Ko\c c University, Rumelifeneri Yolu, 34450 Sar{\i}yer, Istanbul, Turkey. \\
$^2$Department of Physics, Faculty of Science and Letters, Istanbul Technical University, 34469 Maslak, Istanbul, Turkey. 
}
\date{\today}

\begin{abstract}
We use the self-consistent mean-field theory to analyze the effects of Rashba-type 
spin-orbit coupling (SOC) on the ground-state phase diagram of population-imbalanced 
Fermi gases throughout the BCS-BEC evolution. We find that the SOC 
and population imbalance are counteracting, and that this competition tends 
to stabilize the uniform superfluid phase against the phase separation. 
However, we also show that the SOC stabilizes (destabilizes) the uniform 
superfluid phase against the normal phase for low (high) population 
imbalances. In addition, we find topological quantum phase transitions
associated with the appearance of momentum space regions with zero 
quasiparticle energies, and study their signatures in the momentum distribution.
\end{abstract}

\maketitle

\textit{Introduction.}
The recent realization of synthetic gauge fields with neutral bosonic atoms~\cite{nist1}, 
evidently seen from the appearance of vortices in a  BEC, 
has sparked a new wave of theoretical interest in the cold atom community. 
This novel technique uses a spatially-dependent optical coupling between 
internal states of the atoms, and can be used to engineer more complicated
gauge fields by dressing two atomic spin states with a pair of lasers. 
For instance, it has recently been used to create and study the effects of
SOC in a neutral atomic BEC with equal Rashba and Dresselhaus 
strengths~\cite{nist2}. Since this method is equally applicable for neutral 
fermionic atoms, and given that the coupling between a quantum particle's spin 
and its momentum is crucial for the topological insulators and quantum spin 
Hall states, which have recently received immense interest in the condensed 
matter community~\cite{hasan}, it may allow for the realization of topologically 
nontrivial states in atomic systems with possibly a broad interest in 
the physics community~\cite{sato, kubasiak, tewari}. 

Motivated by the recent success in realization of the SOC Bose 
gases~\cite{nist2}, and by a practical proposal for generating a SOC in $^{40}$K atoms~\cite{sau}, 
effects of the SOC have recently been studied for the two-component Fermi 
gases: (i) the two-body problem exactly~\cite{shenoy1}, and (ii) the many-body problem 
in the BCS mean-field approximation~\cite{kubasiak, shenoy2, tewari, zhai, hui}. 
It has been found that the increased density of states 
due to the SOC plays a crucial role for both problems. In particular, for the 
two-body problem, this gives rise to a two-body bound state even on the 
BCS side ($a_s < 0$) of a resonance~\cite{shenoy1}. 
For the many-body problem, the increased density of states favors 
the pairing so significantly that increasing the SOC, while the scattering 
length is held fixed, eventually induces a BCS-BEC crossover even for 
a weakly-interacting system when $a_s \to 0^-$~\cite{shenoy2, zhai, hui}.
In addition, the SOC leads to an anisotropic superfluid, the signatures of
which could be observed in the momentum distribution or the single-particle
spectral function for sufficiently strong SOCs~\cite{hui}.

\begin{figure} [htb]
\centerline{\scalebox{0.65}{\includegraphics{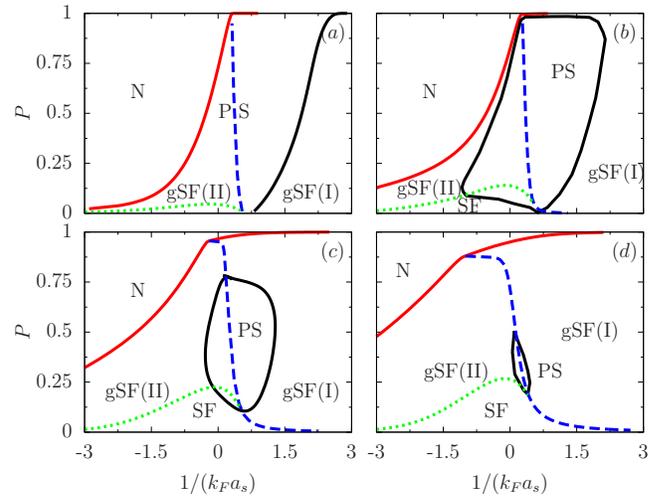}}}
\caption{\label{fig:one} (Color online)
The ground-state phase diagrams are shown as a function of population
imbalance $P = (N_\uparrow - N_\downarrow)/N$ and scattering parameter 
$1/(k_F a_s)$, where the SOC parameter $m \alpha/k_F$ is set to $0.005$ 
in (a), $0.05$ in (b), $0.15$ in (c), and $0.25$ in (d). 
We show normal (N), phase separation (PS), gapped superfluid (SF) and 
gapless superfluid (gSF) phases. 
The dashed blue and dotted green lines separate topologically distinct gSF 
regions, and the trivial SF phase resides at and around the $P = 0$ line.
}
\end{figure}

In this paper, we study the competition between the SOC and population 
imbalance on the ground-state phase diagram of two-component Fermi gases 
across a Feshbach resonance, i.e. throughout the BCS-BEC evolution.
Our main results are shown in Fig.~\ref{fig:one}, and they are as follows.
In the absence of a SOC, the phase diagram of population-imbalanced 
mixtures is well-studied in the literature~\cite{giorgini}, and it mainly 
involves normal (N), phase separation (PS), and topologically distinct 
gapless superfluid (gSF) as well as the trivial gapped superfluid (SF) phases. 
In particular, at and around unitarity ($|a_s| \to \infty$), the system changes from SF 
to PS and then to N as a function of increasing population imbalance.
In this paper, we show that the SOC and population imbalance are 
counteracting, and that this competition always tends to stabilize 
the gSF phase against the PS. 
However, we also show that the SOC stabilizes (destabilizes) 
the gSF phase against the N phase for low (high) population imbalances.
In addition, since the SOC stabilizes the gSF phase for a very large 
parameter region at and around the unitarity, where most experiments 
are conducted~\cite{imbalance}, it may allow for a possible realization
of the gSF phase for the first time with cold atoms.

\textit{Mean-field Hamiltonian.}
To obtain these results, we use the mean-field Hamiltonian~\cite{gorkov, tewari} 
(in units of $\hbar = 1 = k_B$)
\begin{align}
H = \frac{1}{2} \sum_{\mathbf{k}} \psi_\mathbf{k}^\dagger 
 \left( \begin{array}{cccc}
\xi_{\mathbf{k},\uparrow} & S_\mathbf{k} & 0 & \Delta \\
S_\mathbf{k}^* & \xi_{\mathbf{k},\downarrow} & -\Delta & 0  \\
0 & -\Delta^* & -\xi_{\mathbf{-k},\uparrow} & -S_\mathbf{-k}^* \\
\Delta^* & 0 &  -S_\mathbf{-k} & -\xi_{\mathbf{-k},\downarrow}
\end{array} \right)
\psi_\mathbf{k},
\end{align}
which  is defined up to the constant 
$
C = (1/2) \sum_{\mathbf{k},\sigma} \xi_{\mathbf{k},\sigma} + |\Delta|^2/g.
$ 
Here,
$
\psi_{\mathbf{k}}^\dagger = 
[a_{\mathbf{k},\uparrow}^\dagger, a_{\mathbf{k},\downarrow}^\dagger,  a_{\mathbf{-k},\uparrow}, a_{\mathbf{-k},\downarrow}]
$
denotes the fermionic operators collectively, where $a_{\mathbf{k},\sigma}^\dagger$ 
($a_{\mathbf{k},\sigma}$) creates (annihilates) a spin-$\sigma$ fermion with 
momentum $\mathbf{k}$, 
$
\xi_{\mathbf{k},\sigma} = \xi_{\mathbf{-k},\sigma} = \epsilon_{\mathbf{k},\sigma} - \mu_\sigma
$
has the inversion symmetry with $\epsilon_{\mathbf{k},\sigma} = k^2/(2m_\sigma)$ the kinetic energy, 
$\mu_\sigma$ the chemical potential, and $k = \sqrt{k_x^2+k_y^2+k_z^2}$.
Here, $\Delta = g\langle a_{\mathbf{k},\uparrow} a_{-\mathbf{k},\downarrow} \rangle$ is the 
mean-field order parameter, where $g \ge 0$ is the strength of the attractive particle-particle
interaction which is assumed to be local, and $\langle \cdots \rangle$
is a thermal average.
While we keep the formalism quite general including the possibility of 
mass-imbalanced fermion mixtures, we present our  numerical results 
only for mass-balanced mixtures. In addition, we consider only a Rashba-type 
SOC, i.e. $S_\mathbf{k} = -S_{-\mathbf{k}} = \alpha(k_y - ik_x)$, 
where $\alpha \ge 0$ is its strength, but the extension of our work to anisotropic 
or more complicated SOCs is straightforward.

\textit{Self-consistency equations.}
Transforming the mean-field Hamiltonian to the helicity basis, and following the usual 
procedure, i.e. $\partial \Omega / \partial |\Delta| = 0$ for the order parameter and 
$N_\uparrow + s N_\downarrow = - \partial \Omega / \partial \mu_s$ for the number 
equations where $\Omega$ is the mean-field thermodynamic potential, $s = \pm$ 
and $\mu_s = (\mu_\uparrow + s \mu_\downarrow)/2$, we obtain the self-consistency
equations
\begin{align}
\label{eqn:gap}
\frac{2|\Delta|}{g} &= \frac{1}{2} \sum_{\mathbf{k},s} \frac{\partial E_{\mathbf{k},s}}{\partial |\Delta|} \tanh\left( \frac{E_{\mathbf{k},s}}{2T}\right), \\
\label{eqn:ntot}
N_\uparrow + N_\downarrow &= \frac{1}{2} \sum_{\mathbf{k},s} \left[ 1 + \frac{\partial E_{\mathbf{k},s}}{\partial \mu_+} \tanh\left( \frac{E_{\mathbf{k},s}}{2T}\right) \right], \\
\label{eqn:ndiff}
N_\uparrow - N_\downarrow &= \frac{1}{2} \sum_{\mathbf{k},s}  \frac{\partial E_{\mathbf{k},s}}{\partial \mu_-} \tanh\left( \frac{E_{\mathbf{k},s}}{2T}\right).
\end{align}
Here, $T$ is the temperature and
$
E_{\mathbf{k},s}^2 = \xi_{\mathbf{k},+}^2+\xi_{\mathbf{k},-}^2+|\Delta|^2+|S_\mathbf{k}|^2+2 s A_{\mathbf{k}} 
$
gives the quasiparticle excitation spectrum~\cite{kubasiak, tewari}, where
$
A_{\mathbf{k}} = \sqrt{\xi_{\mathbf{k},-}^2(\xi_{\mathbf{k},+}^2 + |\Delta|^2) + |S_\mathbf{k}|^2 \xi_{\mathbf{k},+}^2},
$
$
\xi_{\mathbf{k},s} = \epsilon_{\mathbf{k},s} - \mu_s
$
with
$
\epsilon_{\mathbf{k},s} = (\epsilon_{\mathbf{k},\uparrow} + s\epsilon_{\mathbf{k},\downarrow})/2 = k^2/(2m_s)
$
and
$
m_s = 2m_\uparrow m_\downarrow / (m_\downarrow + s m_\uparrow).
$
Note that $m_+$ is twice the reduced mass of $\uparrow$ and $\downarrow$ particles, and
$m_- \to \infty$ for mass-balanced ($m_\uparrow = m_\downarrow$) mixtures.
In Eqs.~(\ref{eqn:gap}) -~(\ref{eqn:ndiff}), the derivatives of the quasiparticle energies 
are given by
$
\partial E_{\mathbf{k},s} / \partial |\Delta| = (1 + s \xi_{\mathbf{k},-}^2/A_{\mathbf{k}}) |\Delta| / E_{\mathbf{k},s}
$
for the order parameter,
$
\partial E_{\mathbf{k},s} / \partial \mu_+ = - [1 + s( \xi_{\mathbf{k},-}^2+|S_\mathbf{k}|^2)/A_{\mathbf{k}}] \xi_{\mathbf{k},-} / E_{\mathbf{k},s}
$
for the average chemical potential and
$
\partial E_{\mathbf{k},s} / \partial \mu_- = - [1 + s( \xi_{\mathbf{k},+}^2+|\Delta|^2)/A_{\mathbf{k}}] \xi_{\mathbf{k},-} / E_{\mathbf{k},s}
$
for the half of the chemical potential difference.

Equations~(\ref{eqn:gap}) -~(\ref{eqn:ndiff}) are the generalization 
of the mean-field expressions to the case of population- and/or mass-imbalanced mixtures, and
they recover the known expressions (a) when $\xi_{\mathbf{k}, -} = 0$ for which 
$E_{\mathbf{k},s}$ simplifies to
$
E_{\mathbf{k},s}^2 = (\xi_{\mathbf{k},+} + s |S_\mathbf{k}|)^2 + |\Delta|^2,
$ 
and (b) when $|S_{\mathbf{k}}| = 0$ for which $E_{\mathbf{k},s}$ simplifies to
$
E_{\mathbf{k},s}^2 = [s \xi_{\mathbf{k},-} + \sqrt{\xi_{\mathbf{k},+}^2 + |\Delta|^2}]^2.
$
We eliminate the theoretical parameter $g$ in favor of the experimentally
relevant $s$-wave scattering length $a_s$ via the relation,
$
1/g = -m_+ V/(4\pi a_s) + \sum_\mathbf{k} 1/(2\epsilon_{\mathbf{k},+}),
$
where $V$ is the volume. $g$ can also be eliminated in favor of the two-body binding
energy $\epsilon_b \le 0$ in vacuum via the relation~\cite{shenoy1}
$
1/g = (1/2) \sum_{\mathbf{k},s} 1/(2\varepsilon_{\mathbf{k},s}  + \epsilon_{th} - \epsilon_b),
$
where 
$
\varepsilon_{\mathbf{k},s} =  \epsilon_{\mathbf{k},+} + s\sqrt{\epsilon_{\mathbf{k},-}^2+|S_\mathbf{k}|^2} 
$
is the single-particle noninteracting dispersion in the helicity basis, and 
$\epsilon_{th}$ is the threshold for the two-body scattering. 
For the Rashba-type SOC, and assuming $m_\uparrow \le m_\downarrow$, we obtain
$
\epsilon_{th} = 2\alpha^2 m_-(m_- - \sqrt{m_-^2-m_+^2})/m_+,
$
which gives $\epsilon_{th} = m_+\alpha^2$ when $m_- \gg m_+$, i.e. $m_\uparrow \lesssim m_\downarrow$.

\textit{Thermodynamic stability.} 
To construct the phase diagram, we solve the self-consistency equations 
and check the stability of these solutions for the uniform superfluid phase 
using the compressibility (or the curvature) criterion~\cite{iskin, stability}. 
This says that the compressibility matrix $\mathbf{\kappa}(T)$ with elements
$
\kappa_{\sigma,\sigma'} (T) = - \partial^2 \Omega / (\partial \mu_\sigma \partial \mu_{\sigma'})
$
needs to be positive definite, and it is directly related to the condition that the 
curvature of $\Omega$ with respect to $|\Delta|$, i.e.
\begin{align}
&\frac{\partial^2 \Omega}{\partial |\Delta|^2} = \frac{1}{2} \sum_{\mathbf{k},s}  \left\lbrace
- \frac{1}{4T} \left( \frac{\partial E_{\mathbf{k},s}}{\partial |\Delta|} \right)^2 \mathrm{sech}^2\left(\frac{E_{\mathbf{k},s}}{2T}\right) 
 \right. \nonumber \\ 
 &\left. 
 + 
 \left[
   s\frac{|\Delta|^2 \xi_{\mathbf{k},-}^4}{A_{\mathbf{k}}^3} +
   \left( \frac{\partial E_{\mathbf{k},s}}{\partial |\Delta|} \right)^2 
 \right]
 \frac{\tanh\left(\frac{E_{\mathbf{k},s}}{2T}\right)}{2E_{\mathbf{k},s}}
\right\rbrace,
\end{align}
needs to be positive. 
When at least one of the eigenvalues of $\mathbf{\kappa} (T)$, or the curvature 
$\partial^2 \Omega / \partial |\Delta|^2$ is negative, the uniform mean-field solution 
does not correspond to a minimum of $\Omega$, and a nonuniform superfluid phase, 
e.g.  a phase separation, is favored~\cite{iskin, stability}.

\textit{Rashba-type SOC for mass- and population-balanced mixtures.}
To get more insight into the effects of the SOC in the BCS-BEC evolution, let us first 
consider mass- and population-balanced ($m_\sigma = m$ and $N_\sigma = N/2$) 
mixtures, where $N$ is the total number of fermions. This case is analytically more 
tractable, since the single-particle energy in the helicity basis simplifies to~\cite{gorkov}
$
\varepsilon_{\mathbf{k},s} = k^2/(2m) + s \alpha k_\perp,
$
where $k_\perp = \sqrt{k_x^2 + k_y^2}$. For instance, the bound-state equation 
can be solved to obtain
$
1/a_s = \sqrt{m^2 \alpha^2 - m\epsilon_b} + m\alpha\ln[\sqrt{-m\epsilon_b}/(\sqrt{m^2\alpha^2-m\epsilon_b}+m\alpha)].
$
In the weak SOC limit, when $m \alpha^2 \ll \epsilon_b$,  this expression gives
$
\epsilon_b \approx -1/(ma_s^2) + m\alpha^2,
$
up to the leading order in $\alpha$, which recovers the usual result in the $\alpha \to 0$ limit.
However, in the strong SOC limit, when $m \alpha^2 \gg \epsilon_b$, the general 
expression gives
$
\epsilon_b \approx -(4m\alpha^2/e^2) e^{-2/(m\alpha a_s)},
$
which implies that a bound state exists even for $a_s < 0$, although its energy is 
exponentially small~\cite{shenoy1}. This is a result of increased density of states
$
D_s(\epsilon) = \sum_{\mathbf{k}} \delta(\epsilon - \varepsilon_{\mathbf{k},s})
$
due to the SOC~\cite{shenoy1, dos}, where $\delta(x)$ is the Dirac-delta function.

In the case of noninteracting ($g = 0$ or $a_s \to 0^-$) mass- and population-balanced 
Fermi gases at zero temperature ($T = 0$), where $\mu_\sigma = \mu$, calculating the total 
number $N = N_+ + N_- = k_F^3 V / (3\pi^2)$ of fermions, where
$
N_s = \sum_\mathbf{k} \theta(\mu-\varepsilon_{\mathbf{k},s}),
$
we obtain 
$
\mu \approx \epsilon_F - 3m\alpha^2/2
$
up to the leading order in $\alpha$ when $m\alpha^2 \ll \epsilon_F$, and 
$
\mu = - m\alpha^2/2 + 2k_F^3/(3\pi m^2\alpha)
$
when $\mu < 0$. Note that we conveniently choose the energy (length) scale as the 
Fermi energy $\epsilon_F$ (momentum $k_F$) of the $N_\sigma = N/2$ fermions.

It has been shown that increasing the SOC for a noninteracting Fermi 
gas leads to a change in the Fermi surface topology, when the number of fermions 
in the $+$-helicity band ($N_+$) vanishes~\cite{shenoy2}. This occurs when $\mu$ goes 
below the bottom of the energy band, i.e. $\mu = 0$, or when $\alpha$ increases beyond
$
\alpha = [4/(3\pi)]^{1/3} k_F/m \approx 0.75 k_F/m.
$
In some ways, this is similar to the usual BCS-BEC crossover problem, 
where the quasiparticle excitation spectrum changes behavior as a function of increasing
the scattering parameter $1/(k_F a_s)$ at $\mu = 0$, i.e. its minimum is located at a finite (zero) momenta 
when $\mu > 0$ ($\mu < 0$). However, the topological transition discussed here 
is driven by increasing the SOC parameter $\alpha$, and the origin of it can also be 
traced back to the change in the quasiparticle excitation spectrum when $g$ is finite. 
For instance, for mass- and population-balanced mixtures, the excitation spectrum 
simplifies to~\cite{shenoy2, zhai, hui, gorkov}
$
E_{\mathbf{k},s} = \sqrt{(\varepsilon_{\mathbf{k},s} - \mu)^2 + |\Delta|^2},
$
and it also has a change of behavior at $\mu = 0$. 
Having set up the formalism, we are now ready to discuss the competition between 
normal fluidity, uniform superfluidity, and phase separation across a Feshbach resonance.

\textit{Ground-state phase diagrams.}
In the absence of a SOC and at low $T$, it is well-established that the 
self-consistent solutions of Eqs.~(\ref{eqn:gap}) -~(\ref{eqn:ndiff}) are sufficient to 
describe the physics of fermion mixtures both in the BCS and the BEC limits, 
and that these equations also capture qualitatively the correct physics 
in the entire BCS-BEC evolution~\cite{giorgini, iskin}. Hoping that the mean-field 
formalism remains sufficient in the presence of a SOC, here we analyze only the 
ground-state phase diagram of population-imbalanced but mass-balanced
mixtures as a function of both the SOC and scattering parameters. 

There are typically three phases in our phase diagrams. While the normal 
(N) phase is characterized by $\Delta = 0$, the uniform superfluid (SF) 
and nonuniform superfluid, e.g. phase separation (PS), are 
characterized by $\partial^2\Omega / \partial |\Delta|^2 > 0$ and 
$\partial^2\Omega / \partial |\Delta|^2 < 0$, respectively, 
when $\Delta \ne 0$. Furthermore, in addition to the topologically trivial 
gapped SF phase, the gapless SF (gSF) phase can also be distinguished by two 
topologically distinct regions, depending on the momentum-space topology of 
their quasiparticle excitation spectrum (see below).

\begin{figure} [htb]
\centerline{\scalebox{0.65}{\includegraphics{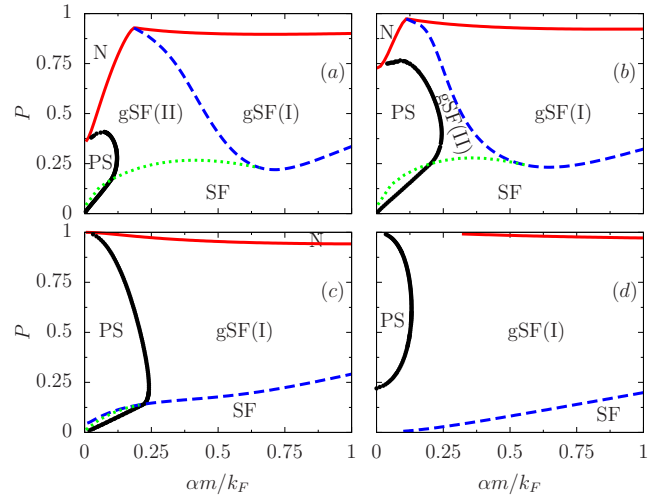}}}
\caption{\label{fig:two} (Color online)
The ground-state phase diagrams are shown as a function of 
$P = (N_\uparrow - N_\downarrow)/N$ and $\alpha$, where $1/(k_F a_s)$ is
set to $-0.5$ in (a),  $0$ in (b),  $0.5$ in (c), and $1.5$ in (d). 
The labels are described both in Fig.~\ref{fig:one} and in the text.
}
\end{figure}

In Fig.~\ref{fig:one}, the phase diagrams are shown as a function of
population imbalance $P = (N_\uparrow - N_\downarrow)/N$ and scattering
parameter $1/(k_F a_s)$ for four different $\alpha$ values.
Comparing these results with the $\alpha \to 0$ limit~\cite{iskin}, it 
is clearly seen that the SOC and population imbalance are counteracting. 
On one hand, this competition always tends to stabilize the gSF phase against 
the PS, and therefore, at any given $P$, the system eventually becomes a 
gSF by increasing $\alpha$, no matter how small $1/(k_F a_s)$ is. 
This is best seen in Fig.~\ref{fig:two}, where the phase diagrams are shown as 
a function of $P$ and $\alpha$ for four different $1/(k_F a_s)$ values.
On the other hand, we find that while the SOC stabilizes the gSF phase 
against the N phase for low $P$ due to increased density of states~\cite{dos}, 
it destabilizes the gSF phase against the N phase for high $P$.

In both figures, the dashed blue and dotted green lines are obtained from the 
conditions $|\Delta|^2 = -\mu_\uparrow \mu_\downarrow$ and $|\Delta| = |\mu_-|$ 
(see below), respectively, and they separate 
the topologically distinct gSF regions. The trivial gapped SF phase resides 
at and around the $P = 0$ lines. As can be inferred from 
Fig.~\ref{fig:two}, the dashed blue line makes a dip as $1/(k_F a_s) \lesssim 0$, 
the tip of which eventually touches the $P = 0$ line at $\alpha \approx 0.75k_F/m$, 
consistent with our analysis above for the topological transition of a noninteracting 
($a_s \to 0^-$) system. In Figs.~\ref{fig:one} and~\ref{fig:two}, we also 
show that the SOC stabilizes the gSF for a very large parameter region, 
at and around the unitarity, which is normally unstable against PS when 
$\alpha = 0$, allowing for a possible realization with cold atoms for the first time, 
as discussed next in great details.

\textit{Topological phase transition.}
Population imbalance is achieved when either $E_{\mathbf{k},+}$ 
(for $N_\downarrow > N_\uparrow$) or $E_{\mathbf{k},-}$ (for $N_\uparrow > N_\downarrow$) 
has zeros in some places of $\mathbf{k}$-space. Let us assume $N_\uparrow \ge N_\downarrow$
without loosing generality, for which $E_{\mathbf{k},+}$ is always gapped.
Depending on the number of zeros of $E_{\mathbf{k},-}$ (zero energy 
surfaces in $\mathbf{k}$-space), there are two topologically distinct gSF phases:
gSF(I) where $E_{\mathbf{k},-}$ has two, and gSF(II) where $E_{\mathbf{k},-}$ 
has four zeros. 
The zeros of $E_{\mathbf{k},-}$ can be found by imposing the condition
$
E_{\mathbf{k},+}^2 E_{\mathbf{k},-}^2 = (\xi_{\mathbf{k},\uparrow} \xi_{\mathbf{k},\downarrow} 
+ |\Delta|^2 - |S_\mathbf{k}|^2)^2+4|\Delta|^2 |S_\mathbf{k}|^2 = 0,
$
indicating that both $|S_\mathbf{k}| = 0$ and 
$\xi_{\mathbf{k},\uparrow} \xi_{\mathbf{k},\downarrow} + |\Delta|^2 = 0$
needs to be satisfied. 

For the Rashba-type SOC that we consider in this paper, the zeros occur when 
$k_\perp = 0$ and at real $k_z$ momenta,
$
k_{z,s}^2 = B_+ + s \sqrt{B_-^2 - 4m_\uparrow m_\downarrow |\Delta|^2},
$
provided that 
$
|\Delta|^2 < |B_-|^2/(4m_\uparrow m_\downarrow)
$
for $B_+ \ge 0$, and 
$
|\Delta|^2 < - \mu_\uparrow \mu_\downarrow
$
for $B_+ < 0$. Here,
$
B_s = m_\uparrow \mu_\uparrow + s m_\downarrow \mu_\downarrow.
$
Note that the conditions on $k_z$ coincide with those obtained for $k$ 
in the case of population-imbalanced mixtures without the SOC where 
$E_{\mathbf{k},s}$ is isotropic.
The topologically trivial SF phase corresponds to the case where both 
$E_{\mathbf{k},+}$ and $E_{\mathbf{k},-}$ have no zeros and are always gapped.
The transition from gSF(II) to gSF(I) occurs when $|k_{z,-}| \to 0$,
indicating a change in topology in the lowest quasiparticle band, 
similar to the Lifshitz transition in ordinary metals and nodal 
(non-$s$-wave) superfluids. 

\begin{figure} [htb]
\centerline{\scalebox{0.6}{\includegraphics{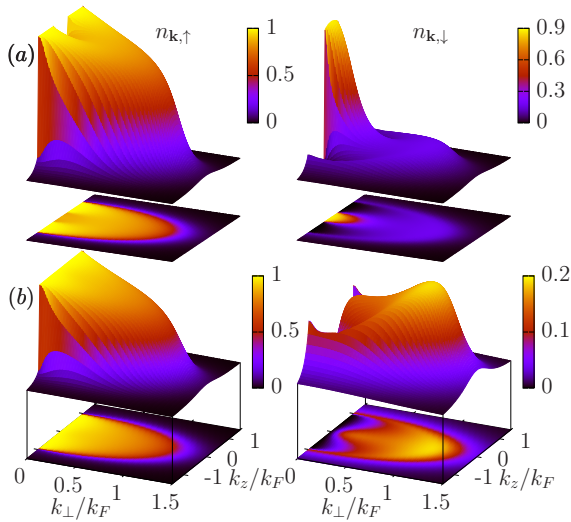}}}
\caption{\label{fig:three} (Color online)
Typical momentum distributions $n_{\mathbf{k},\sigma}$ 
are shown as a function of $k_\perp = \sqrt{k_x^2 + k_y^2}$ and $k_z$, where
we set $P = 0.5$ and $1/(k_F a_s) = 0$, and vary the SOC parameter:
(a) $\alpha = 0.275 k_F/m$ (for which $|\Delta| = 0.463\epsilon_F$, 
$\mu_+ = 0.550\epsilon_F$ and $\mu_- = 0.653\epsilon_F$),
and (b) $\alpha = 0.350 k_F/m$ (for which $|\Delta| = 0.465\epsilon_F$, 
$\mu_+ = 0.427\epsilon_F$ and $\mu_- = 0.688\epsilon_F$),
corresponding to gSF(II) and gSF(I) phases, respectively.
}
\end{figure}

The topological transition here is unique, because it involves an $s$-wave 
superfluid, and could be potentially observed for the first time through the 
measurement of the momentum distributions $n_{\mathbf{k},\sigma}$ of $\uparrow$
and $\downarrow$ fermions, both of which are readily available from
Eqs.~(\ref{eqn:ntot}) and~(\ref{eqn:ndiff}). For instance, we illustrate the typical 
$T = 0$ distribution of a gSF(II) phase in Fig.~\ref{fig:three}(a), and of a 
gSF(I) phase in Fig.~\ref{fig:three}(b). In these figures, we note 
that $n_{\mathbf{k},\sigma}$ is anisotropic in $\mathbf{k}$ space, 
which follows from the anisotropic structure of $E_{\mathbf{k},s}$. For 
$\mathbf{k}$-space regions where $k_\perp = 0$ and $k_{z,-} \le |k_z| \le k_{z,+}$, the 
corresponding distributions are exactly $n_{\mathbf{k},\uparrow} = 1$ and 
$n_{\mathbf{k},\downarrow} = 0$. Here, $k_{z,s}$ are approximately found to 
be $k_{z,-} = 0.30k_F$ and $k_{z,+} = 1.00k_F$ in Fig.~\ref{fig:three}(a), and 
$k_{z,-} = 0$ and $k_{z,+} = 0.97k_F$ in Fig.~\ref{fig:three}(b), in perfect agreement
with our analysis above. We also see that a major redistribution occurs for the 
minority species ($n_{\mathbf{k},\downarrow}$) at the gSF(II) to gSF(I) transition 
boundary, where the sharp peak that is present near the origin vanishes abrubtly.
Although this topological transition is quantum ($T=0$) in nature, signatures of 
it should still be observed at finite $T$ within the quantum critical region, 
where the $n_{\mathbf{k},\sigma}$ are smeared out due to thermal effects. 
Although the primary signature of this topological transition is seen 
in $n_{\mathbf{k},\sigma}$, single-particle spectral function~\cite{rf} as well as 
some thermodynamic quantities such as the atomic compressibility would also 
show an anomaly at the transition boundary.

\textit{Conclusions.}
In summary, we analyzed the effects of SOC on the ground-state phase 
diagram of population-imbalanced Fermi gases throughout the BCS-BEC 
evolution. We found that the SOC and population imbalance are 
counteracting, and that this competition always tends to stabilize the gSF 
phase against the PS. In contrast, while the SOC stabilizes the gSF phase 
against the N phase for low population imbalances, it destabilizes the gSF 
phase against the N phase for high population imbalances. 
In addition, we found topological quantum phase transitions associated with 
the appearance of momentum space regions with zero quasiparticle energies, 
and studied their signatures in the momentum distribution. 
We hope that our work will motivate further research in this direction, 
since the SOC stabilizes the gSF phase for a very large parameter region at 
and around the unitarity, allowing for a possible realization of the gSF phase 
for the first time with cold atoms.

\textit{Acknowledgments.}
This work is supported by the Marie Curie International Reintegration 
(Grant No. FP7-PEOPLE-IRG-2010-268239), Scientific and Technological 
Research Council of Turkey (Career Grant No. T\"{U}B$\dot{\mathrm{I}}$TAK-3501-110T839), 
and the Turkish Academy of Sciences (T\"{U}BA-GEB$\dot{\mathrm{I}}$P award).


\begin{thebibliography}{99}
\bibitem{dos}  For a Rashba-type SOC, when $m_\sigma = m$, we obtain
$
D_s(\epsilon \ge 0) = [mV/(2\pi^2)] [ \sqrt{2m\epsilon} + m\alpha \pi (1-s)/2 - m\alpha \arcsin\left( \sqrt{\frac{2m\epsilon}{2m\epsilon + m^2\alpha^2}}  \right) ] 
$
and
$
D_s(\epsilon < 0) = [mV (1-s)/(4\pi^2)]  ( \sqrt{2m\epsilon + m^2\alpha^2} + m\alpha\pi/2) \theta(m\alpha^2+2\epsilon),
$
where $\theta(x)$ is the heaviside-step function.
\end{thebibliography}
\end{document}